\documentclass[prd,aps,superscriptaddress,preprintnumbers,notitlepage,nofootinbib]{revtex4}
\usepackage{tablefootnote}
\usepackage[T1]{fontenc}
\usepackage{amsmath}
\usepackage{amssymb}
\usepackage{adjustbox}
\usepackage{multirow}
\usepackage{hyperref}
\usepackage{bbm}
\usepackage{xcolor}
\usepackage{braket}
\usepackage{slashed}
\usepackage{diagbox}
\usepackage[utf8]{inputenc}
\usepackage{tcolorbox}
\usepackage{tabularx}
\usepackage{bm}

\newcommand{\eqal}[1]{\begin{align}#1\end{align}}

\newcommand{\nb}{\nonumber}

\newcommand{\p}[1]{$\mathbf{P}$}
\newcommand{\cp}[1]{$\mathbf{CP}$}

\begin{document}
\raggedbottom

\title{Parity Violation on Longitudinal Single-Spin Asymmetries at the EicC}

\author{Yong Du}\email{yongdu5@impcas.ac.cn}
\affiliation{Tsung-Dao Lee Institute, Shanghai Jiao Tong University, Shanghai 200240, China} 
\affiliation{Institute of Modern Physics, Chinese Academy of Sciences, Lanzhou 730000, China} 
\affiliation{School of Nuclear Science and Technology, University of Chinese Academy of Sciences,\\
19A Yuquan Road, Beijing 100049, China} 
\date{\today}

\begin{abstract}
We explore two longitudinal single-spin asymmetries induced from parity violation in neutral-current deep inelastic scattering at the proposed Electron-ion collider in China (EicC): $A_{PV}^{e\,(p)}$ from longitudinally polarized (unpolarized) electrons scattering off unpolarized (longitudinally polarized) protons. We find $A_{PV}^e$, of $\mathcal{O}(10^{-4})$, is generically one to three orders of magnitude larger than $A_{PV}^p$. We further estimate different uncertainty sources including statistics, parton distribution functions, and beam polarization, for both asymmetries, and then identify individually their dominance in different regimes of the Bjorken-$x$. Based on these results, we then advocate utilizing $A_{PV}^p$ for the extraction of the weak mixing angle at two representative momentum transfer scales unexplored before, and we find a relative precision below 10\% can be achieved at the EicC with an effective one-year operation time.
\end{abstract}

\maketitle

\section{Introduction}\label{sec:intro}
From the first proposal of parity violation\,\cite{Lee:1956qn,Wu:1957my,Garwin:1957hc,Friedman:1957mz} to the first evidence of the weak neutral current from the Gargamelle experiment at CERN in 1973\,\cite{GargamelleNeutrino:1973jyy}, parity-violating (PV) electron scattering had played an important role in confirming the gauge-structure of the Standard Model (SM) and ruling out alternative theories\,\cite{Cahn:1977uu,Prescott:1978tm,Prescott:1979dh}.\footnote{Recently, it was also shown that parity violation can have a non-trivial impact on quantum entanglement and Bell nonlocality\,\cite{Du:2024sly}.} Accordingly, the PV asymmetry, denoted as $A_{PV}$, can be measured by flipping the beam polarization of the incident beam(s) and also provides a measure of one of the key electroweak parameters being the weak mixing angle $\sin\theta_w$. Many experiments have been carried out in this respect, see\,\cite{ParticleDataGroup:2024cfk} for a comprehensive review and summary, including as well the electron-ion collider (EIC) commissioned by the BNL and the JLab\,\cite{Boer:2011fh,Accardi:2012qut}, the MOLLER experiment at the JLab\,\cite{MOLLER:2014iki}, and the P2 experiment at Mainz\,\cite{Berger:2015aaa} in the near future. Note that recently the two-loop theoretical computation of $A_{PV}^{\rm MOLLER}$ was finished in\,\cite{Du:2019evk}, as well as the recent studies on $A_{PV}$ at the $J/\psi$ threshold in\,\cite{Bondar:2019zgm,Fu:2023ose,Du:2024jfc}.

\begin{figure}[!t]
\includegraphics[width = 0.48 \linewidth]{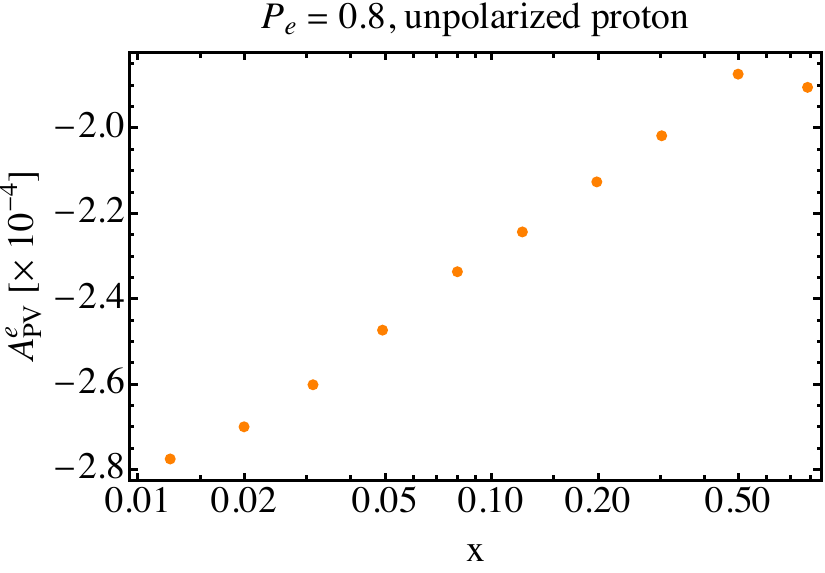}
\includegraphics[width = 0.48 \linewidth]{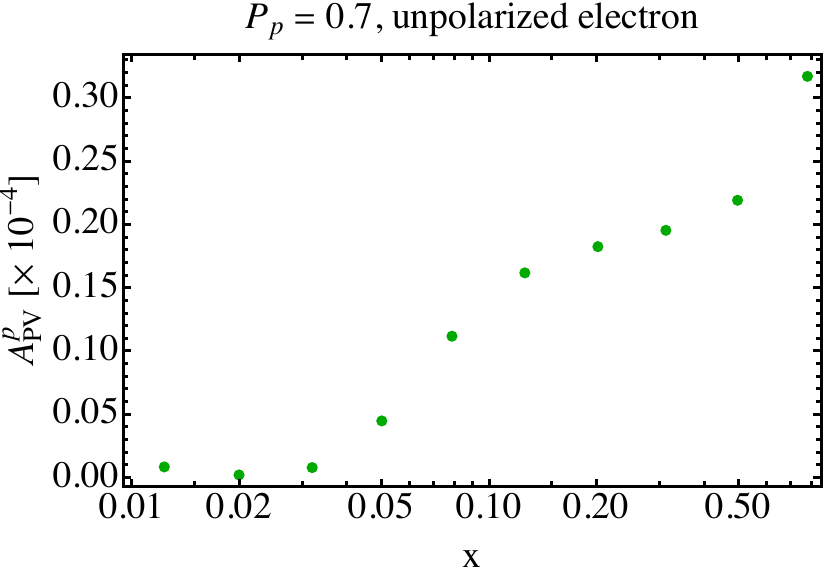}
\caption{
\textit{Single-spin} asymmetries as a function of the Bjorken-$x$. See the main text for details.
} \label{fig:apvep}
\end{figure}

In this work, we focus on the PV deep inelastic scattering at the proposed electron-ion collider in China (EicC). This proposed program will be based on the High-Intensity heavy-ion Accelerator Facility (HIAF) in Huizhou, providing polarized electron and/or proton beams with a center-of-mass energy ranging from 15\,GeV to 20\,GeV. See\,\cite{Anderle:2021wcy} for a detailed description of this project. The uniqueness of EIC and EicC is that they both provide a chance to measure $A_{PV}$ at energy scales unexplored before, rendering them also promising for precision SM tests similarly to the other PV electron scattering experiments discussed in the last paragraph. For recent studies on the EIC or deep inelastic scattering experiments, see for example, \cite{Liu:2002bq,Diaconescu:2004aa,Ramsey-Musolf:2005qlt,Zhao:2016rfu,Cirigliano:2021img,Yan:2021htf,Liu:2021lan,Li:2021uww,Yan:2022npz,Boughezal:2022pmb,Bacchetta:2023hlw,Balkin:2023gya,Wang:2024zns,Wen:2024cfu,Deng:2025hio} and references therein. In this study, we focus uniquely on the EicC program operating at a relatively lower energy scale than the EIC and concentrate on the following two scenarios: (1) electron PV asymmetry $A_{PV}^e$ from longitudinally polarized electrons with 80\% beam polarization scattering off unpolarized protons, and (2) proton PV asymmetry $A_{PV}^p$ from unpolarized electron scattering off longitudinally polarized protons with 70\% beam polarization. For each scenario, we investigate the sensitivity reach of EicC to the longitudinal single-spin asymmetries (defined below) and estimate their uncertainties in order to obtain their prospects at the EicC. We find that:
\begin{itemize}
\item $A_{PV}^e$ is generically larger than $A_{PV}^p$ by one to three orders of magnitude, as seen in our figure\,\ref{fig:apvep};
\item the statistical uncertainty is negligible for $A_{PV}^e$, and similarly for $A_{PV}^p$ but only for $x\gtrsim0.1$. The dominant uncertainty in $A_{PV}^e$ is thus from the unpolarized structure function $F_3^{\gamma Z}$, with that from $F_2^{\gamma\gamma,\gamma Z}$ and electron beam polarization subdominant and comparable. In contrast, the statistical error and that from the polarized structure function $g_{1,5}^{\gamma Z}$ dominates at $x\lesssim0.1$ for $A_{PV}^p$. See our figure\,\ref{fig:staterror};
\item The total uncertainty of $A_{PV}^e$ is generically much larger than that of $A_{PV}^p$ as summarized in figure\,\ref{fig:apvepfin}, and we expect both their errors to be reduced with the upcoming datasets available at the EIC and the EicC in the near future. Furthermore, the relatively smaller uncertainty in $A_{PV}^p$ also renders it promising for a precision determination of the weak mixing angle below 10\% at the EicC, as depicted in figure\,\ref{fig:angle}.
\end{itemize}

To obtain these results, we organize the rest of this work as follows: in section\,\ref{sec:theo}, we set up the theoretical framework for constructing and evaluating the longitudinal single-spin asymmetries. Then in section\,\ref{sec:res} we present our results, where we detail our simulations for computing the two asymmetries, and how we quantify different sources of uncertainties such as statistics, parton distribution functions, and beam polarization. By quantifying the dominance of different uncertainty sources in different Bjorken-$x$ regimes, we then discuss which longitudinal single-spin asymmetry to utilize for the extraction of the weak mixing angle. We then conclude in section\,\ref{sec:conclu}.

\section{Theoretical Setup}\label{sec:theo}
As mentioned in the introduction, we will only consider electron scattering off a proton in this work via neutral-current exchange and denote the process generically as $e^-(k,s_1) + p (p,s_p) \to e^-(k',s_2) + X(p_X,s_X)$. Here, $X$ represents the remnant of the proton after the scattering, and $s_i$ the spin of $i$. The total amplitude is parameterized as follows
\eqal{
i\mathcal{M} = &\, \sum_{V'=\gamma,Z}\bar u(k',s_2)ig_{V'ee}\gamma_\alpha (g_V^{V',e} - g_A^{V',e}\gamma_5) u(k,s_1) \frac{-i}{q^2-m_V'^2} ig_{V'pp}M_{V'}^\alpha (p,p_X,s_p,s_X),
}
with $q=k-k'$ the momentum transfer, $g_{V,A}^{V',e}$ the vector ($V$) and axial-vector ($A$) couplings between $V'$ and the electron normalized by $g_{V'ee}$, where $g_{Zee} = g_L/(2c_w)$, $g_V^{Z,e}=T_3^e-2Q_e s_w^2$, $g_A^{Z,e}=T_3^e$, and $g_{\gamma ee} = Q_e e$, $g_V^{\gamma,e}=1$, $g_A^{\gamma,e}=0$. Here, $T_3^f$ and $Q_f$ are the isospin and electric charge of $f$ in units of the proton charge, respectively. $M_{V'}^\alpha(p,p_X,s_p,s_X)$ represents the hadronic current depending on the momenta and spins of $p$ and $X$, which is normalized similarly as the SM currents with $g_{\gamma pp}=e Q_p$, $g_{Z pp}=g_L/(2c_w)$.

For EicC, available options include either polarized or unpolarized electron and proton beams. As a consequence, the inclusive invariant amplitude can be obtained by summing over the spins of final state particles to obtain:
\eqal{
\langle \mathcal{M}^2 \rangle = &\, \frac{e^4}{(q^2)^2} L_{\gamma\gamma}^{\alpha\beta}(s_1) [M_{\gamma\gamma}]_{\alpha\beta}(s_p) + \frac{g_L^4}{16c_w^4} \frac{1}{(q^2-m_Z^2)^2} L_{ZZ}^{\alpha\beta}(s_1) [M_{ZZ}]_{\alpha\beta}(s_p)\nb\\
&\, - \frac{e^2g_L^2}{4c_w^2} \frac{1}{q^2(q^2-m_Z^2)} L_{\gamma Z}^{\alpha\beta}(s_1) [M_{\gamma Z}]_{\alpha\beta}(s_p),
}
where the first and second terms represent pure $\gamma$ and $Z$ contributions, and the last term that from their interference. $L^{\alpha\beta}_{V_1V_2}$ is the leptonic tensor with the polarization of the final state electron summed over and
\eqal{
L_{\gamma\gamma}^{\alpha\beta}(\lambda_e) = &\, 2\left( k^\alpha k'^{\beta} + k^\beta k'^{\alpha} - k\cdot k' g^{\alpha\beta} + i \lambda_e \epsilon^{\alpha\beta\mu\nu}k'_\mu k_{\nu} \right),\\
L_{\gamma Z}^{\alpha\beta}(\lambda_e) = &\, \left( g_V^e - \lambda_e g_A^e \right)L_{\gamma\gamma}^{\alpha\beta}(s_1),\quad L_{Z Z}^{\alpha\beta}(\lambda_e) = \left( g_V^e - \lambda_e g_A^e \right)^2 L_{\gamma\gamma}^{\alpha\beta}(\lambda_e),
}
where $\lambda_e$ is the helicity of the electron and terms suppressed by the electron mass $m_e$ have been discarded. The differential rate can then be computed directly by integrating over the phase space of the final state particles, for which we opt to use the following broadly adopted Lorentz invariants defined as
\eqal{
\nu = &\, \frac{q\cdot p}{m_p}, \quad Q^2 = -q^2, \quad x = \frac{Q^2}{2m_p \nu}, \quad y = \frac{q\cdot p}{k\cdot p}, \quad W^2 = (p+q)^2, \quad s = (k+p)^2.
}
The differential rate can then be expressed as
\eqal{
\frac{d^2\sigma}{dxdy} = &\, \frac{2\pi y \alpha^2}{Q^4} L_{\gamma\gamma}^{\alpha\beta} \left[ \biggr( \eta_{\gamma\gamma} W^{\gamma\gamma}_{\alpha\beta} + \eta_{\gamma Z} \left( g_V^e - \lambda_e g_A^e \right) W^{\gamma Z}_{\alpha\beta} + \eta_{ZZ} \left( g_V^e - \lambda_e g_A^e \right)^2 W^{ZZ}_{\alpha\beta} \biggr) \right],
}
where again we ignore terms suppressed by $m_e$ and define
\eqal{
W^{V_1V_2}_{\alpha\beta} &\, = \frac{1}{4\pi}\int \frac{d^3p_X}{(2\pi)^3 2E_{\bm p_X}} (2\pi)^4\delta^{(4)}(p+k-k'-p_X) [M_{V_1V_2}]_{\alpha\beta}(s_p),\\
\eta_{\gamma \gamma} &\, = 1, \quad \eta_{\gamma Z} = - \frac{G_F m_Z^2}{2\sqrt2\pi\alpha_{\rm EM}} \frac{Q^2}{(Q^2 + m_Z^2)}, \quad \eta_{ZZ} = \eta_{\gamma Z}^2.
}
Here, $G_F$ is the Fermi constant as determined from the muon lifetime, and $\alpha_{\rm EM}$ is the fine structure constant.

For the hadronic tensors $W_{\alpha\beta}^{V_1V_2}$, they can be generically parameterized as\,\cite{Blumlein:1996vs}\footnote{See\,\cite{Blumlein:1998nv} for accounting for target mass corrections.}
\eqal{
W_{\mu \nu}^{V_1V_2} = & \left(-g_{\mu \nu}+\frac{q_\mu q_\nu}{q^2}\right) F_1^{V_1V_2}\left(x, Q^2\right)+\frac{\hat{p}_\mu \hat{p}_\nu}{p \cdot q} F_2^{V_1V_2}\left(x, Q^2\right) - i \varepsilon_{\mu \nu \alpha \beta} \frac{q^\alpha p^\beta}{2 p \cdot q} F_3^{V_1V_2}\left(x, Q^2\right) \nb \\
& + i \varepsilon_{\mu \nu \alpha \beta} \frac{q^\alpha}{p \cdot q}\left[s^\beta g_1^{V_1V_2}\left(x, Q^2\right)+\left(s^\beta-\frac{s \cdot q}{p \cdot q} p^\beta\right) g_2^{V_1V_2}\left(x, Q^2\right)\right] \nb \\
&\, + \frac{1}{p \cdot q}\left[\frac{1}{2}\left(\hat{p}_\mu \hat{s}_\nu+\hat{s}_\mu \hat{p}_\nu\right)-\frac{s \cdot q}{p \cdot q} \hat{p}_\mu \hat{p}_\nu\right] g_3^{V_1V_2}\left(x, Q^2\right) \nb\\
&\, + \frac{s \cdot q}{p \cdot q}\left[\frac{\hat{p}_\mu \hat{p}_\nu}{p \cdot q} g_4^{V_1V_2}\left(x, Q^2\right)+\left(-g_{\mu \nu}+\frac{q_\mu q_\nu}{q^2}\right) g_5^{V_1V_2}\left(x, Q^2\right)\right],
}
where
\eqal{
\hat{p}_\mu=p_\mu-\frac{p \cdot q}{q^2} q_\mu, \quad \hat{s}_\mu=s_\mu-\frac{s \cdot q}{q^2} q_\mu,
}
and $s^\mu\equiv s_p^\mu$ is the four-spin of the proton as a reflection of its polarization.\footnote{If the proton is unpolarized instead, one then needs to sum over the helicity states of the proton as we will discuss shortly below.} The structure functions are evaluated using the quark-parton model\,\cite{ParticleDataGroup:2024cfk} in this work, which predicts
\eqal{
{\left[F_2^{\gamma\gamma}, F_2^{\gamma Z}, F_2^{ZZ} \right] } &\, = 2x {\left[F_1^{\gamma\gamma}, F_1^{\gamma Z}, F_1^{ZZ} \right] } = x \sum_f\left[Q_f^2, 2 Q_f g_V^f, \left(g_V^{f}\right)^2+\left(g_A^{f}\right)^2 \right]\left(q_f+\bar{q}_f\right), \\ 
{\left[F_3^{\gamma\gamma}, F_3^{\gamma Z}, F_3^{ZZ} \right] } &\, = \sum_f\left[0,2 Q_f g_A^f, 2 g_V^f g_A^f \right]\left(q_f-\bar{q}_f\right), \\ 
{\left[g_1^{\gamma\gamma}, g_1^{\gamma Z}, g_1^{ZZ} \right] } &\, = \frac{1}{2} \sum_f\left[Q_f^2, 2 Q_f g_V^f, \left(g_V^{f}\right)^2+\left(g_A^{f}\right)^2 \right]\left(\Delta q_f+\Delta \bar{q}_f\right), \\ 
{\left[g_4^{\gamma\gamma}, g_4^{\gamma Z}, g_4^{ZZ} \right] } &\, = 2x {\left[g_5^{\gamma\gamma}, g_5^{\gamma Z}, g_5^{ZZ} \right] } = 2x \sum_f\left[0, Q_f g_A^f, g_V^f g_A^f \right]\left(\Delta q_f-\Delta \bar{q}_f\right),
}
with $q_f$ and $\Delta q_f$ the unpolarized and polarized parton distribution distribution (PDF) of flavor $f$ in the proton and the Callan-Gross relations explicitly indicated in the first and the fourth lines above. The differential cross section can then be written compactly as
\eqal{
\frac{d^2\sigma}{dxdy}\left(\lambda_e,\lambda_p\right) = &\, \frac{4\pi \alpha^2}{xyQ^2} \biggr[ xy^2 F_1 + (1-y) F_2 + \frac{\lambda_e}{2} xy (2-y) F_3 - \lambda_e \lambda_p xy (2-y) g_1 + \lambda_p (1-y) g_4 + \lambda_p xy^2 g_5 \biggr],
}
with $\lambda_p$ the helicity of the proton and 
\eqal{
\left[F_{1,2,3}, \, g_{1,4,5}\right] &\, = \sum\limits_{V_1V_2={\gamma\gamma,\gamma Z,ZZ}} \eta_{V_1V_2} \xi_{V_1V_2} \left[F_{1,2,3}, \, g_{1,4,5}\right]^{V_1V_2}(x,Q^2),\\
\left[ \xi_{\gamma\gamma}, \, \xi_{\gamma Z}, \, \xi_{ZZ} \right] &\, = \left[ 1, \left( g_V^e - \lambda_e g_A^e \right), \left( g_V^e - \lambda_e g_A^e \right)^2 \right].
}
Note that only the $F_{1,2,3}$ and $g_{1,4,5}$ structure functions survive since the other ones are suppressed by $m_{e,p}^2/Q^2$.


Depending on the polarization of the electron and/or the proton, several asymmetries can be constructed to extract different structure functions. We are particularly interested in the parity-violating ones in this study and will consider only longitudinally polarized electron and/or proton beams in the following though transverse polarization is also a viable option of EicC. The differential cross section in this case can then be classified according to the helicity of the beams:
\eqal{
\frac{d^2\sigma^{\rm long.}}{dxdy} = &\, \frac14\biggr[ (1+P_e)(1+P_p)\left.\frac{d\sigma}{dxdy}\right\vert_{\lambda_{e,p}=+1} + (1-P_e)(1-P_p)\left.\frac{d\sigma}{dxdy}\right\vert_{\lambda_{e,p}=-1} \nb\\
&\, \qquad + (1+P_e)(1-P_p)\left.\frac{d\sigma}{dxdy}\right\vert_{\lambda_{e}=-\lambda_{p}=+1} + (1-P_e)(1+P_p)\left.\frac{d\sigma}{dxdy}\right\vert_{\lambda_{p}=-\lambda_{e}=+1} \biggr]\nb\\
\equiv &\, \frac{d\sigma_0}{dxdy} + P_e \frac{d\sigma_e}{dxdy} + P_p \frac{d\sigma_p}{dxdy} + P_e P_p \frac{d\sigma_{ep}}{dxdy},
}
where
\eqal{
\frac{d^2\sigma_0}{dxdy} &\, = \frac14 \sum_{\lambda_{e,p}=\pm1} \frac{d^2\sigma}{dxdy}\left(\lambda_e,\lambda_p\right), \quad \frac{d^2\sigma_e}{dxdy} = \frac14 \sum_{\lambda_{p}=\pm1} {\rm sgn}(\lambda_e) \frac{d^2\sigma}{dxdy}\left(\lambda_e,\lambda_p\right), \\
\frac{d^2\sigma_p}{dxdy} &\, = \frac14 \sum_{\lambda_{e}=\pm1} {\rm sgn}(\lambda_p) \frac{d^2\sigma}{dxdy}\left(\lambda_e, \lambda_p \right), \quad \frac{d^2\sigma_{ep}}{dxdy} = \frac14 \sum_{\lambda_{e,p}=\pm1} \lambda_e\lambda_p \frac{d^2\sigma}{dxdy}\left(\lambda_e, \lambda_p\right),
}
and ``sgn'' represents the sign function. To connect with experiments, we consider asymmetric observables that can be directly related to the observed number of events in the following:
\begin{itemize}
\item the \textit{single-spin} asymmetry where either the proton or the electron is polarized, but not both of them. In this case, we define
\eqal{
A_{PV}^e &\, \equiv \left\vert P_e\right\vert \frac{d^2\sigma_e }{d^2\sigma_0} = \frac{ \left\vert P_e\right\vert \eta_{\gamma Z}\left[2 (y-1) g_A^e F_2^{\gamma Z} - x y \left( 2 g_A^e y F_1^{\gamma Z} - (2-y) g_V^e F_3^{\gamma Z} \right) \right] }{ 2 \eta_{\gamma\gamma} \left( (1-y) F_2^{\gamma\gamma} + x y^2 F_1^{\gamma\gamma} \right)  },\\
A_{PV}^p &\, \equiv  \left\vert P_p\right\vert \frac{d^2\sigma_p}{d^2\sigma_0} = \frac{ \left\vert P_p\right\vert \eta_{\gamma Z}\left[- 2 (y-1) g_V^e g_4^{\gamma Z} + 2 x y \left( g_V^e y g_5^{\gamma Z} + (2-y) g_A^e g_1^{\gamma Z} \right) \right] }{ 2 \eta_{\gamma\gamma} \left( (1-y) F_2^{\gamma\gamma} + x y^2 F_1^{\gamma\gamma} \right) },
}
which we call the electron and the proton PV asymmetries $A_{PV}^e$ and $A_{PV}^p$, respectively, in the following. Clearly, the electron beam polarization opens the window for extracting the parity violating structure function $F_3^{\gamma Z}$, and $A_{PV}^p$ to $g_{1,4,5}^{\gamma Z}$. The parity and CP even structure functions $F_{1,2}^{\gamma\gamma}$ can, for example, be determined from the total cross section at different energy runs.
\item the \textit{double-spin} asymmetry where both the electron and the proton are polarized is similarly defined as
\eqal{
A_{PV}^{ep} &\, \equiv \left\vert P_e P_p\right\vert \frac{d^2\sigma_{ep}}{d^2\sigma_0}\nb\\
&\, = \frac{ 2\eta_{\gamma Z} \left\vert P_e P_p\right\vert \left[ g_A^e (y-1) g_4^{\gamma Z} + x y \left( g_V^e (y-2) g_1^{\gamma Z} - g_A^e y g_5^{\gamma Z} \right) \right] + 2 \eta_{\gamma\gamma} \left\vert P_e P_p\right\vert x y (y-2) g_1^{\gamma\gamma} }{ \eta_{\gamma Z}\left[- 2 (y-1) g_V^e F_2^{\gamma Z} + x y \left( 2 g_V^e y F_1^{\gamma Z} - (2-y) g_A^e F_3^{\gamma Z} \right) \right] + 2 \eta_{\gamma\gamma} \left( (1-y) F_2^{\gamma\gamma} + x y^2 F_1^{\gamma\gamma} \right) }.
}
Clearly, $A_{PV}^{ep}$ can be utilized to extract the parity-conserving polarized structure function $g_1^{\gamma\gamma}$, and it conserves parity at the leading order where $\eta_{\gamma Z}\to0$. While itself is an interesting observable, we will not consider this \textit{double-spin} asymmetry and other alternatives in this work from the consideration of parity violation presence at the leading order. 
\end{itemize}
We comment that in both the \textit{single-} and the \textit{double-spin} cases above, we have consistently ignored pure $Z$ contributions given the fact that the momentum transfer at EicC is much smaller than the weak scale. For the same reason, we also leave out the interference in the denominator of the asymmetries defined above except for $A_{PV}^{ep}$. This now naturally leads to the question: How sensitive is EicC to these \textit{single-spin} asymmetries?

\section{Numerical Results}\label{sec:res}
To answer the question at the end of last section, we adopt the proposed setup of EicC as outlined in the white paper\,\cite{Anderle:2021wcy}, with longitudinal electron polarization $P_e = (80\pm1.6)\%$ and longitudinal proton polarization $P_p = (70\pm3.5)\%$. The beam energy of the electron is $E_e = 3.5\rm\,GeV$ and that for the proton is $E_p = 20\rm\,GeV$, corresponding to a center of mass energy $\sqrt s = 16.7$\,GeV. For numerical evaluations, we use {\tt NNPDFpol1.1}\,\cite{Nocera:2014gqa} with 100 replicas for polarized protons as obtained by combining data from CERN, SLAC, DESY, and RHIC. On the other hand, for unpolarized protons, the {\tt MMHT2014}\,\cite{Harland-Lang:2014zoa} PDF with 50 eigenvector sets in total is implemented.

For $A_{PV}^{e,p}$ evaluations, we firstly scan over $x\in[0.01,0.99]$ by dividing it into ten logarithmically spaced bins, and similarly for the momentum transfer with $Q^2\in[1,\,30]\rm\,GeV^2$\,\cite{Anderle:2021wcy}. To present our results, we then evaluate $A_{PV}^{e,p}$ in each $x$ bin by averaging over randomly generated $10^4$ $Q^2$'s with a weight function $w(Q^2)\equiv G_F Q^2/(2\sqrt2\alpha_{\rm EM})\approx \eta_{\gamma Z}$\,\cite{Zhao:2016rfu}, where the weighted $Q^2$ in each $x$ bin is used as the benchmark to obtain the input for the weak mixing angle. The results are shown in figure\,\ref{fig:apvep} for $A_{PV}^e$ (left panel) and $A_{PV}^p$ (right panel) in terms of the Bjorken-$x$. From the plots, one observes that $A_{PV}^e$ is generically of $\mathcal{O}(10^{-4})$ and its magnitude decreases with increasing $x$. This is in contrast to $A_{PV}^p$, whose magnitude increase instead with increasing $x$. Furthermore, we find that $A_{PV}^e$ is generically larger than $A_{PV}^p$ by one to three orders of magnitude, suggesting a potential advantage of the former for experimental measurements provided their uncertainties are comparable.

\begin{figure}[!t]
\includegraphics[width = 0.52 \linewidth]{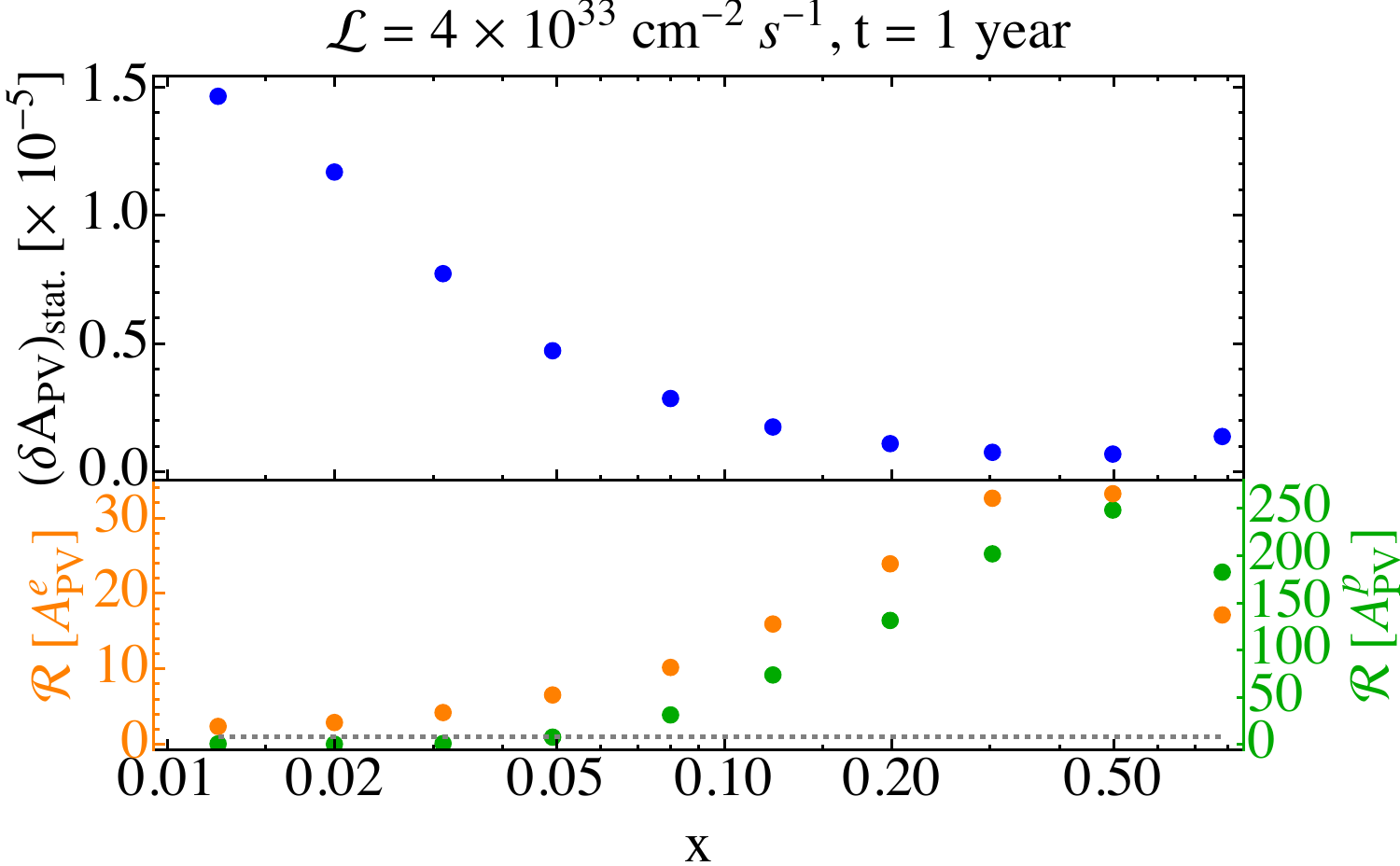}
\includegraphics[width = 0.45 \linewidth]{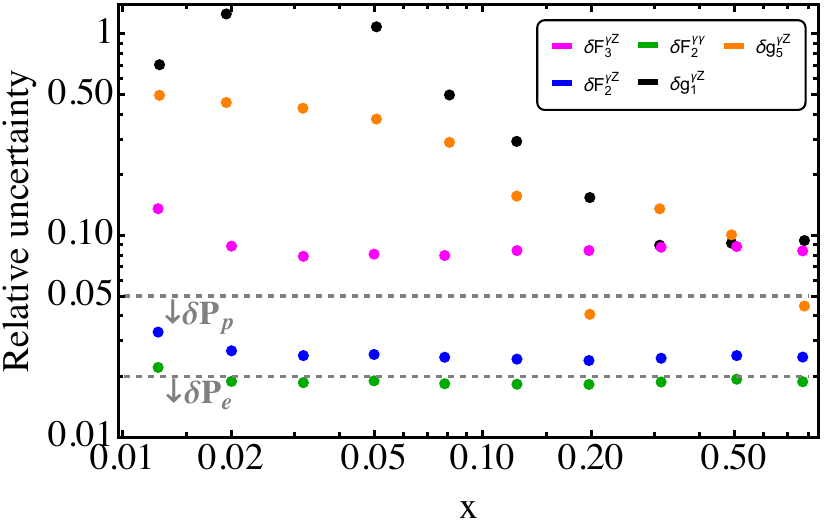}
\caption{
{\bf Left Panel}: Statistical uncertainties $(\delta A_{PV})_{\rm stat.}$ in each $x$ bin are obtained by assuming a conservative detector efficiency of 10\% and effectively one-year operation time and shown in the upper half. The lower half gives the ratio $\mathcal{R}$ between the absolute values of the asymmetries and the statistical errors with $\mathcal{R}=1$ for the horizontal gray dashed line; {\bf Right Panel}: Relative uncertainties for the structure functions as indicated by the legend in the upper right corner. The two horizontal gray dashed lines correspond to the uncertainty in beam polarization for the EicC. See the main text for details.
} \label{fig:staterror}
\end{figure}

In order to quantify the uncertainties in $A_{PV}^{e,p}$, we firstly investigate the statistical one, which we denote as $(\delta A_{PV})_{\rm stat.}$. Due to the smallness of the asymmetries in both cases, one can show that the statistical uncertainty is approximately $1/\sqrt{N_{\rm events}}$, with $N_{\rm events}$ the total number of events from \textit{unpolarized} electron beam scattering off \textit{unpolarized} protons in the specific bin under consideration. To estimate $N_{\rm events}$ in each $x$ bin, we assume a universal but rather conservative detector efficiency of 10\%\,\cite{Anderle:2021wcy}, and then show our results in the left panel of figure\,\ref{fig:staterror} with a luminosity of $4\times10^{33} {\rm\, cm}^{-2} \cdot s^{-1}$\,\cite{Anderle:2021wcy} and an effective one-year operation time. We find from the plot that, though the statistical uncertainty is generically small, it can reach $\mathcal{O}(10^{-5})$ at small $x$ as seen from the upper half of the left panel.\footnote{We have also checked that the total statistical uncertainty agrees with the estimation in\,\cite{Anderle:2021wcy}.} Nevertheless, the statistical error is found significantly smaller than the absolute magnitude of $A_{PV}^e$ as seen from the orange points in the lower half of the left panel of figure\,\ref{fig:staterror}, where we define the $y$ axis as
\eqal{
\mathcal{R} [A_{PV}^{e,p}] \equiv \left\vert \frac{A_{PV}^{e,p}}{(\delta A_{PV})_{\rm stat.}} \right\vert.
}
Clearly, the high luminosity of the EicC renders statistical errors negligible for the determination of $A_{PV}^e$, which in turn helps improve the precision of the unpolarized PDF in the regime we are interested in in this work. However, this is not generically true for $A_{PV}^p$ as shown by the green dots in the same plot, where its $y$ axis in green on the right-hand side shall be understood accordingly. One can see that the statistical uncertainty also becomes negligible for large $x\gtrsim0.1$ while non-negligible otherwise.

Assuming the systematical error will be well under control at the EicC, another source of uncertainty comes from the structure functions in each $x$ bin. In the unpolarized case, this is estimated by assuming symmetric errors in the PDF for simplicity and is then evaluated as\,\cite{Martin:2009iq}
\eqal{
\sigma_f(x,|Q|) = \frac12\sqrt{\sum_i \left( x\, q_f^{2i-1}(x,|Q|) - x\, q_f^{2i}(x,|Q|) \right)^2 },\label{eq:symerr}
}
with $|Q|=\sqrt{Q^2}$ and the summation over the 50 eigenvector sets of {\tt MMHT2014}. In the polarized case, this is evaluated directly using the {\tt NNPDF} setup at the 95\% confidence level (CL). For the former, we find it practically very CPU expensive to evaluate eq.\,\eqref{eq:symerr} in each $x$ bin containing $10^4$ randomly generated $|Q|$'s. We therefore firstly generate $\sigma_f(x,|Q|)$ over a $201\times201$ grid on the $x-|Q|$ plane and then interpolate it for quantifying the uncertainties in these structure functions at the next stage of our simulation with {\tt Mathematica}.

\begin{figure}[!t]
\includegraphics[width = 0.48 \linewidth]{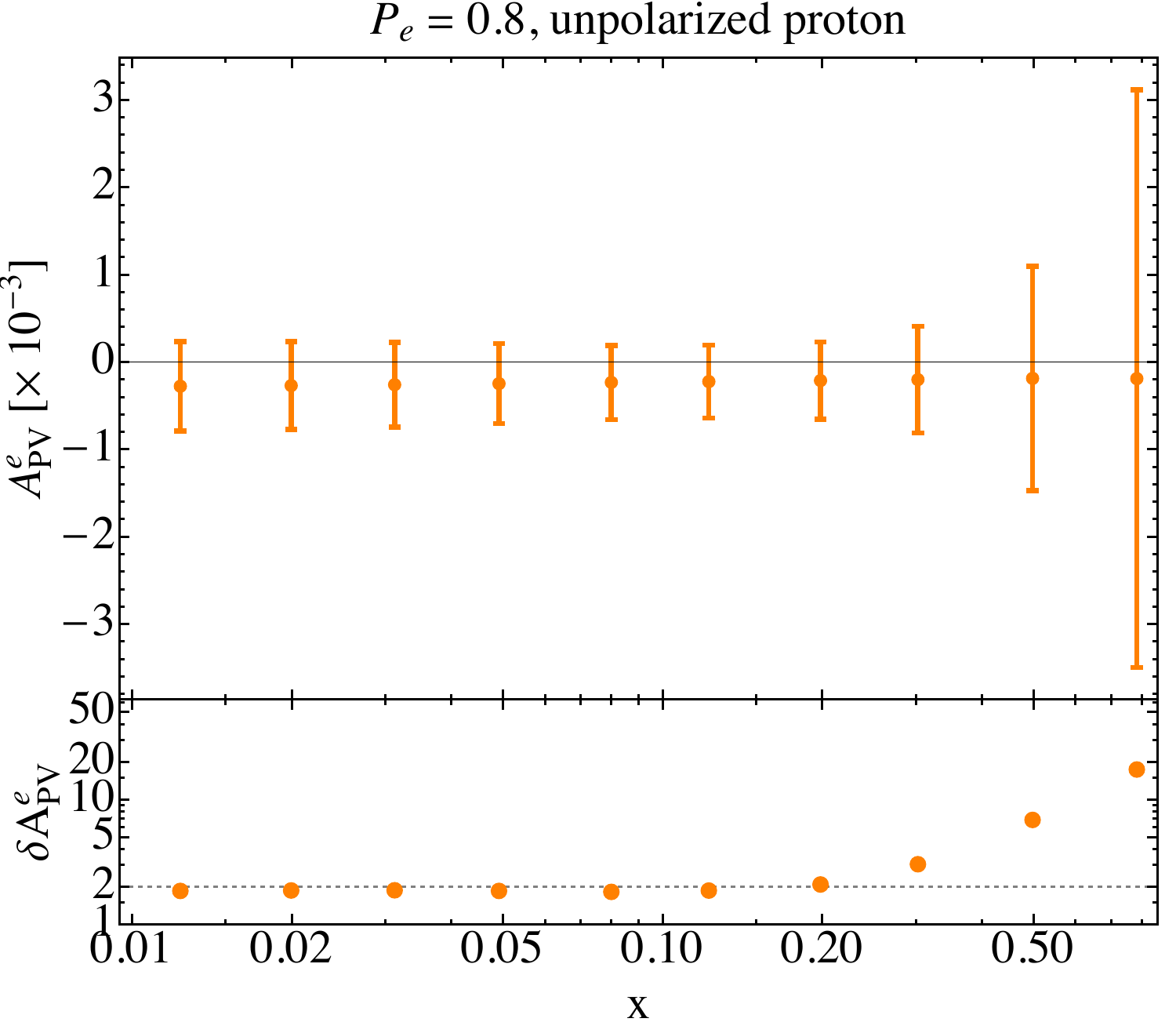}
\includegraphics[width = 0.48 \linewidth]{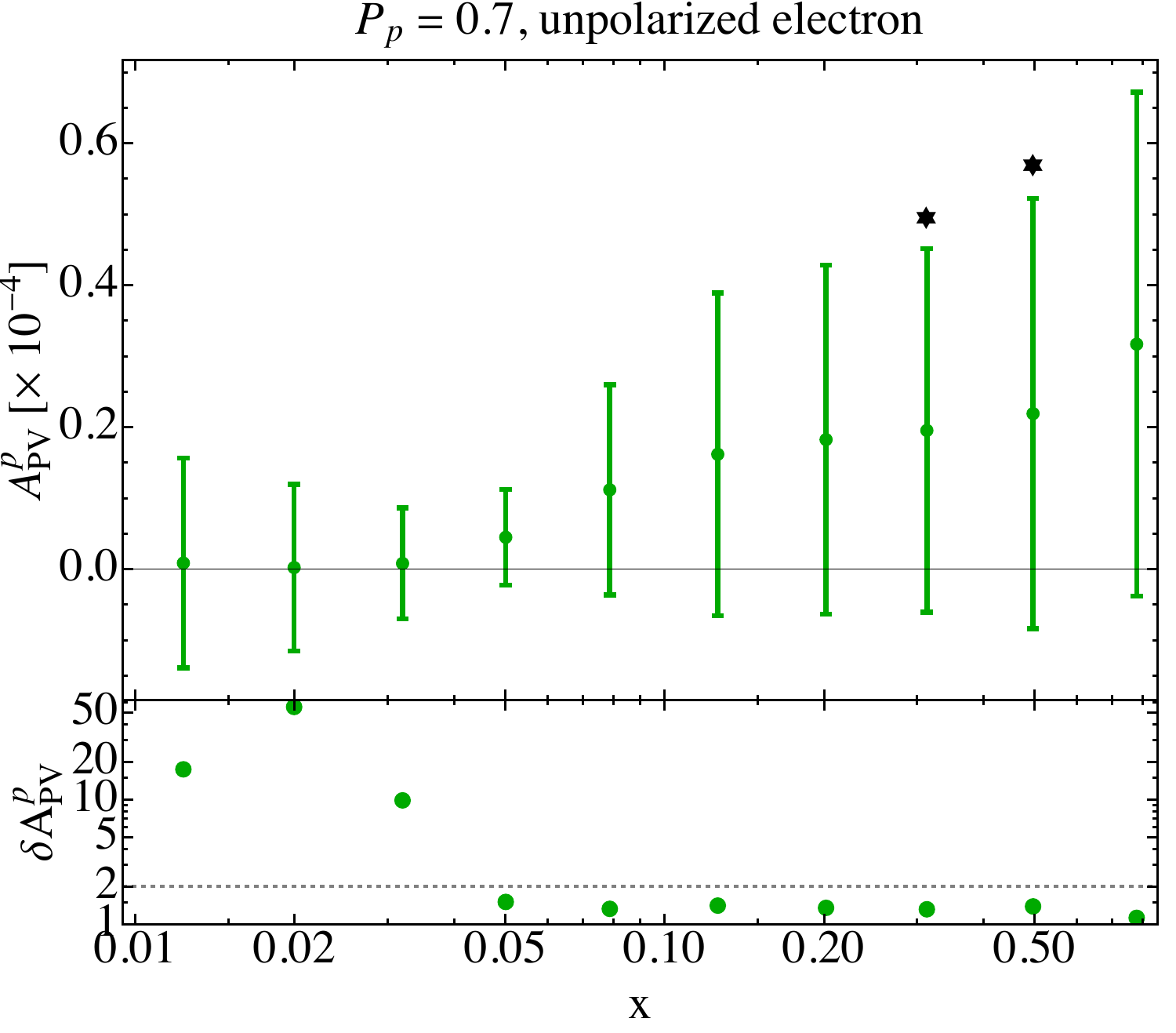}
\caption{
{\bf Left Panel}: Predicted $A_{PV}^{e}$ in each $x$ bin after combining all the errors at the 95\%CL, with its relative uncertainty shown at the bottom; {\bf Right Panel}: Same as the left but for $A_{PV}^{p}$. The two bins with black stars atop correspond to the ones we select for the extraction of the weak mixing angle discussion as detailed in the main text.
} \label{fig:apvepfin}
\end{figure}

The uncertainties in these structure functions are then shown in the right panel of figure\,\ref{fig:staterror} with the vertical axis being the relative uncertainty for each structure function as indicated by the legends. The upper (lower) horizontal gray dashed line corresponds to the relative uncertainty in the proton (electron) beam polarization $P_p$ ($P_e$), below which the uncertainty originated from the beam polarization becomes significant as indicated by the arrows in front. Recall that the statistical uncertainty in $A_{PV}^e$ is always negligible as discussed above, we thus conclude that the dominant error in $A_{PV}^e$ will be from the parity-violating unpolarized structure function $F_3^{\gamma Z}$, as depicted by the magenta points in the right panel of figure\,\ref{fig:staterror}. On the other hand, since the statistical uncertainty in $A_{PV}^p$ also becomes negligible for $x\gtrsim0.1$, the main obstacles in precision $A_{PV}^p$ measurements will thus be from $g_{1,5}^{\gamma Z}$ in this regime. For $x\lesssim0.1$, both the statistical uncertainty and those from $g_{1,5}^{\gamma Z}$ will become important, which have to be reduced, for example, from an improved fit on the PDFs with a richer data sample available at the EicC and the EIC in the near future.


Combining all the uncertainties discussed above, we show the predicted PV asymmetries in the left and right panel of figure\,\ref{fig:apvepfin} for $A_{PV}^e$ and $A_{PV}^p$, respectively. At the bottom of each subplot, we also show accordingly the relative uncertainties of the asymmetries in each $x$ bin. From figure\,\ref{fig:apvepfin}, we see that the relative error in $A_{PV}^e$ \textit{increases} rapidly for $x\gtrsim0.2$. In contrast, that in $A_{PV}^p$ instead \textit{decreases} dramatically for $x\gtrsim0.05$. This can be understood from the fact that in their respective regimes with large uncertainties, the absolute magnitudes of both asymmetries become substantially tiny as already seen in figure\,\ref{fig:apvep}. Therefore, though the absolute magnitude of $A_{PV}^e$ is generically much larger than $A_{PV}^p$ by one to three orders of magnitude as we commented earlier, the much larger uncertainty over the full $x$ range in the former renders it challenging for a definite observation of a non-vanishing asymmetry at the EicC based on the current fitted unpolarized PDF used in this work. This situation may be changed with the high luminosity of the EicC, as well as the EIC, where the PDF uncertainties are expected to be further reduced. Different from the polarized electron case, though the proton PV asymmetry $A_{PV}^p$ is small, we observe that its relative error at $x\gtrsim0.05$ is small, and therefore promote this scenario for future exploration at the EicC especially given the upcoming uncertainty reduction in polarized PDF in the future.

Finally, we comment on the determination of the weak mixing angle using the PV asymmetries studied above. This is particularly interesting for the EicC and the EIC as they provide a chance for measuring $\sin\theta_w$ at scales unexplored before. For the EIC, this has been investigated, for example, in\,\cite{Zhao:2016rfu,Boughezal:2022pmb}. In the following, we address this point for the EicC by focusing on the two $x$ bins with black stars in the right panel of figure\,\ref{fig:apvepfin}. We comment that a similar analysis is feasible utilizing $A_{PV}^e$ instead in principle, but due to its large uncertainties as discussed above, we will only focus on $A_{PV}^p$ in the two selected bins in the following.

\begin{figure}[!t]
\includegraphics[width = 0.55 \linewidth]{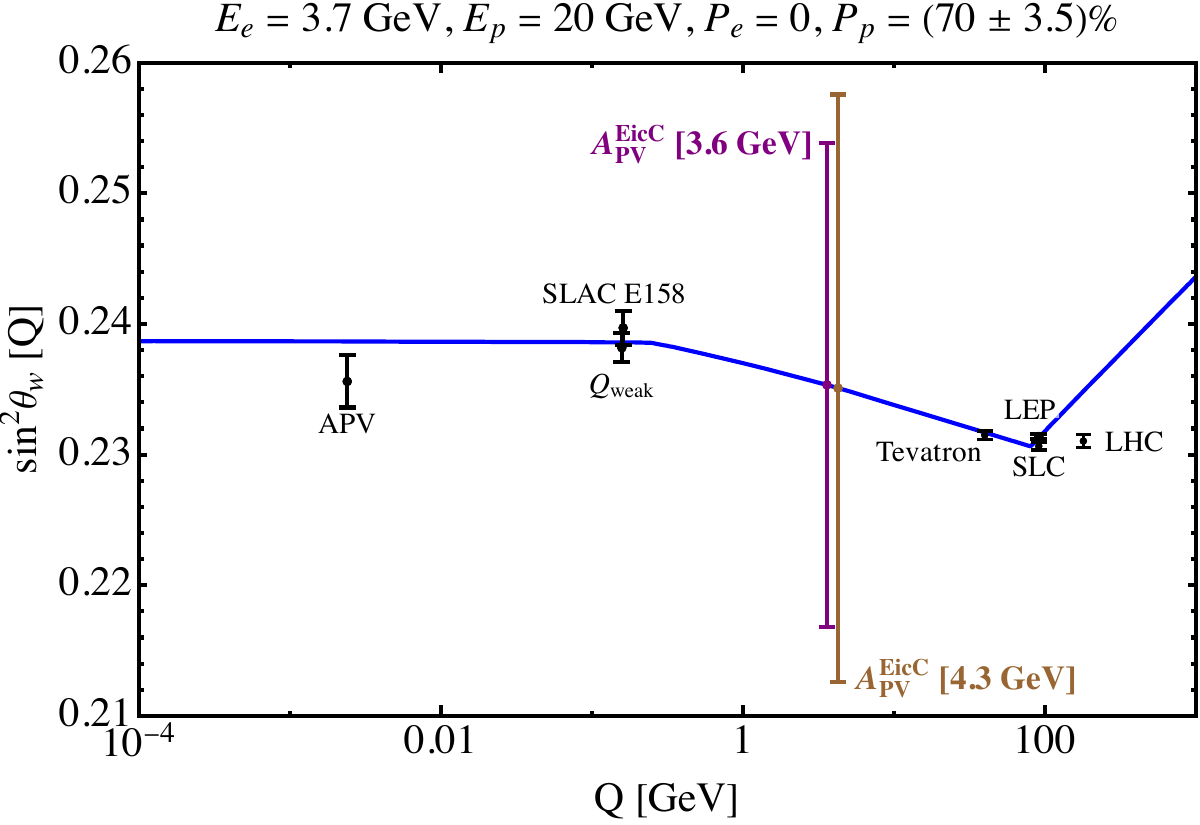}
\caption{
Precision reach of the EicC in measuring the weak mixing angle $\sin^2\theta_w$ from $A_{PV}^p$ at two representative energies $|Q|=3.6\rm\,GeV$ and $|Q|=4.3\rm\,GeV$. The blue curve represents the theoretical prediction obtained in\,\cite{Erler:2004in}.
} \label{fig:angle}
\end{figure}

At each of the two bins, the central value of $\sin^2\theta_w$ is computed by using the weighted $Q^2$ explained above. Specifically, we have $|Q|=3.6\rm\,GeV$ and $|Q|=4.3\rm\,GeV$ in the two bins, respectively. The corresponding $\sin^2\theta_w$ are then evaluated using the results in\,\cite{Erler:2004in}. The uncertainty in each bin is estimated through the standard error propagation, based on which we obtain
\eqal{
&\, [\sin^2\theta_w](3.6\rm\,GeV)_{\rm EicC} = 0.235\pm 0.019,\\
&\, [\sin^2\theta_w](4.3\rm\,GeV)_{\rm EicC} = 0.235\pm 0.022,
}
corresponding to a precision below 10\%. These results are also pictorially shown in figure\,\ref{fig:angle} for comparison with the other existing results. It is worth stressing again that the uncertainties in $\sin^2\theta_w$ above are obtained using the current polarized PDF {\tt NNPDFpol1.1}, which are expected to be reduced when more datasets from (near) future experiments such as the EicC and the EIC become available.

\section{Conclusions}\label{sec:conclu}
The proposed EicC experiment in Huizhou, China is expected to deliver longitudinally polarized electrons with polarization $P_e = (80\pm1.6)\%$ and/or polarized proton beams with polarization $P_p = (70\pm3.5)\%$ for deep inelastic scattering studies. Its luminosity is expected to reach $4\times 10^{33}\,{\rm cm}^{-2}\,s^{-1}$ with a center of mass energy of 16.7\,GeV for $e$-$p$ scattering and a momentum coverage of $1\,{\rm GeV}^2\lesssim Q^2 \lesssim 30\,{\rm GeV}^2$. This makes the EicC a testbed for precision SM studies in a region unexplored before.

In this work, by making use of longitudinally polarized electrons scattering off unpolarized protons, as well as unpolarized electrons scattering off longitudinally polarized protons, we investigate the sensitivity reach of the EicC to single-spin asymmetries $A_{PV}^{e,p}$ as a consequence of parity violation. By comparing different sources of uncertainties as summarized in figure\,\ref{fig:staterror}, we find the statistical uncertainty in $A_{PV}^e$ can always be neglected within the full range of $x$, and it is the unpolarized parity-violating structure function $F_3^{\gamma Z}$ that dominates the total error. In addition, the uncertainty from the electron beam polarization is comparable to that from $F_{2}^{\gamma Z}$ and $F_{2}^{\gamma \gamma}$, and is a factor of a few smaller than that from $F_{3}^{\gamma Z}$. In contrast, for $A_{PV}^p$, the statistical uncertainty becomes negligible for large $x\gtrsim0.1$, in which region the uncertainties from the polarized structure functions $g_{1,5}^{\gamma Z}$ are comparable and are about 10\%. On the other hand, for $x\lesssim0.1$, the statistical uncertainty in $A_{PV}^p$ becomes large and comparable to those from $g_{1,5}^{\gamma Z}$. The full uncertainties for $A_{PV}^{e,p}$ in different $x$ bins are summarized in figure\,\ref{fig:apvepfin}, and we conclude that it will be relatively more challenging to measure $A_{PV}^{e}$ than $A_{PV}^{p}$ since the former has much larger total uncertainties. This conclusion, however, may change when more datasets become available in the near future from the EicC and the EIC when the uncertainties in the PDFs may be further reduced.

Utilizing $A_{PV}^{p}$ from the consideration of its smaller uncertainties, we also perform a determination of the weak mixing angle in two representative bins with $|Q|=3.6$\,GeV and $|Q|=4.3$\,GeV, respectively. We find the EicC can achieve a relative precision of 7.5\% and 9.6\% in measuring $\sin^2\theta_w$ with an effective operation time of one year. This is also pictorially illustrated in figure\,\ref{fig:angle} for comparing with the other existing measurements.

Finally, we comment that since the EicC can reach $Q^2\simeq1\rm\,GeV^2$, it then becomes not well-justified in ignoring terms proportional to $m_p^2/Q^2$ when computing the differential rates. On the other hand, resurrecting these terms will also introduce the dependence on the other polarized structure functions such as $g_{3}^{\gamma\gamma,\gamma Z}$. The high luminosity and low momentum transfer coverage of the EicC may thus provide a unique chance to extract them and further investigate their impact on the parity symmetry.

\section*{Acknowledgements}
YD would like to thank De-Xu Lin for his invitation to visit IMP-CAS and for introducing the EicC to him in 2023, based on which this project was later conceived. YD also thanks DXL for the hospitality during his stay in Lanzhou. This work was partially supported by NSFC Special Funds for Theoretical Physics under grant number 12347116, and partially by the Postdoctoral Fellowship Program of CPSF under number GZC20231613. YD also acknowledges support from the Shanghai Super Postdoc Incentive Plan, and the T.D. Lee Postdoctoral Fellowship at the Tsung-Dao Lee Institute, Shanghai Jiao Tong University.

\bibliography{ref}

\begin{thebibliography}{40}
\expandafter\ifx\csname natexlab\endcsname\relax\def\natexlab#1{#1}\fi
\expandafter\ifx\csname bibnamefont\endcsname\relax
  \def\bibnamefont#1{#1}\fi
\expandafter\ifx\csname bibfnamefont\endcsname\relax
  \def\bibfnamefont#1{#1}\fi
\expandafter\ifx\csname citenamefont\endcsname\relax
  \def\citenamefont#1{#1}\fi
\expandafter\ifx\csname url\endcsname\relax
  \def\url#1{\texttt{#1}}\fi
\expandafter\ifx\csname urlprefix\endcsname\relax\def\urlprefix{URL }\fi
\providecommand{\bibinfo}[2]{#2}
\providecommand{\eprint}[2][]{\url{#2}}

\bibitem[{\citenamefont{Lee and Yang}(1956)}]{Lee:1956qn}
\bibinfo{author}{\bibfnamefont{T.~D.} \bibnamefont{Lee}} \bibnamefont{and}
  \bibinfo{author}{\bibfnamefont{C.-N.} \bibnamefont{Yang}},
  \bibinfo{journal}{Phys. Rev.} \textbf{\bibinfo{volume}{104}},
  \bibinfo{pages}{254} (\bibinfo{year}{1956}).

\bibitem[{\citenamefont{Wu et~al.}(1957)\citenamefont{Wu, Ambler, Hayward,
  Hoppes, and Hudson}}]{Wu:1957my}
\bibinfo{author}{\bibfnamefont{C.~S.} \bibnamefont{Wu}},
  \bibinfo{author}{\bibfnamefont{E.}~\bibnamefont{Ambler}},
  \bibinfo{author}{\bibfnamefont{R.~W.} \bibnamefont{Hayward}},
  \bibinfo{author}{\bibfnamefont{D.~D.} \bibnamefont{Hoppes}},
  \bibnamefont{and} \bibinfo{author}{\bibfnamefont{R.~P.}
  \bibnamefont{Hudson}}, \bibinfo{journal}{Phys. Rev.}
  \textbf{\bibinfo{volume}{105}}, \bibinfo{pages}{1413} (\bibinfo{year}{1957}).

\bibitem[{\citenamefont{Garwin et~al.}(1957)\citenamefont{Garwin, Lederman, and
  Weinrich}}]{Garwin:1957hc}
\bibinfo{author}{\bibfnamefont{R.~L.} \bibnamefont{Garwin}},
  \bibinfo{author}{\bibfnamefont{L.~M.} \bibnamefont{Lederman}},
  \bibnamefont{and} \bibinfo{author}{\bibfnamefont{M.}~\bibnamefont{Weinrich}},
  \bibinfo{journal}{Phys. Rev.} \textbf{\bibinfo{volume}{105}},
  \bibinfo{pages}{1415} (\bibinfo{year}{1957}).

\bibitem[{\citenamefont{Friedman and Telegdi}(1957)}]{Friedman:1957mz}
\bibinfo{author}{\bibfnamefont{J.~I.} \bibnamefont{Friedman}} \bibnamefont{and}
  \bibinfo{author}{\bibfnamefont{V.~L.} \bibnamefont{Telegdi}},
  \bibinfo{journal}{Phys. Rev.} \textbf{\bibinfo{volume}{106}},
  \bibinfo{pages}{1290} (\bibinfo{year}{1957}).

\bibitem[{\citenamefont{Hasert et~al.}(1973)}]{GargamelleNeutrino:1973jyy}
\bibinfo{author}{\bibfnamefont{F.~J.} \bibnamefont{Hasert}}
  \bibnamefont{et~al.} (\bibinfo{collaboration}{Gargamelle Neutrino}),
  \bibinfo{journal}{Phys. Lett. B} \textbf{\bibinfo{volume}{46}},
  \bibinfo{pages}{138} (\bibinfo{year}{1973}).

\bibitem[{\citenamefont{Cahn and Gilman}(1978)}]{Cahn:1977uu}
\bibinfo{author}{\bibfnamefont{R.~N.} \bibnamefont{Cahn}} \bibnamefont{and}
  \bibinfo{author}{\bibfnamefont{F.~J.} \bibnamefont{Gilman}},
  \bibinfo{journal}{Phys. Rev. D} \textbf{\bibinfo{volume}{17}},
  \bibinfo{pages}{1313} (\bibinfo{year}{1978}).

\bibitem[{\citenamefont{Prescott et~al.}(1978)}]{Prescott:1978tm}
\bibinfo{author}{\bibfnamefont{C.~Y.} \bibnamefont{Prescott}}
  \bibnamefont{et~al.}, \bibinfo{journal}{Phys. Lett. B}
  \textbf{\bibinfo{volume}{77}}, \bibinfo{pages}{347} (\bibinfo{year}{1978}).

\bibitem[{\citenamefont{Prescott et~al.}(1979)}]{Prescott:1979dh}
\bibinfo{author}{\bibfnamefont{C.~Y.} \bibnamefont{Prescott}}
  \bibnamefont{et~al.}, \bibinfo{journal}{Phys. Lett. B}
  \textbf{\bibinfo{volume}{84}}, \bibinfo{pages}{524} (\bibinfo{year}{1979}).

\bibitem[{\citenamefont{Du et~al.}(2024{\natexlab{a}})\citenamefont{Du, He,
  Liu, and Ma}}]{Du:2024sly}
\bibinfo{author}{\bibfnamefont{Y.}~\bibnamefont{Du}},
  \bibinfo{author}{\bibfnamefont{X.-G.} \bibnamefont{He}},
  \bibinfo{author}{\bibfnamefont{C.-W.} \bibnamefont{Liu}}, \bibnamefont{and}
  \bibinfo{author}{\bibfnamefont{J.-P.} \bibnamefont{Ma}}
  (\bibinfo{year}{2024}{\natexlab{a}}), \eprint{2409.15418}.

\bibitem[{\citenamefont{Navas et~al.}(2024)}]{ParticleDataGroup:2024cfk}
\bibinfo{author}{\bibfnamefont{S.}~\bibnamefont{Navas}} \bibnamefont{et~al.}
  (\bibinfo{collaboration}{Particle Data Group}), \bibinfo{journal}{Phys. Rev.
  D} \textbf{\bibinfo{volume}{110}}, \bibinfo{pages}{030001}
  (\bibinfo{year}{2024}).

\bibitem[{\citenamefont{Boer et~al.}(2011)}]{Boer:2011fh}
\bibinfo{author}{\bibfnamefont{D.}~\bibnamefont{Boer}} \bibnamefont{et~al.}
  (\bibinfo{year}{2011}), \eprint{1108.1713}.

\bibitem[{\citenamefont{Accardi et~al.}(2016)}]{Accardi:2012qut}
\bibinfo{author}{\bibfnamefont{A.}~\bibnamefont{Accardi}} \bibnamefont{et~al.},
  \bibinfo{journal}{Eur. Phys. J. A} \textbf{\bibinfo{volume}{52}},
  \bibinfo{pages}{268} (\bibinfo{year}{2016}), \eprint{1212.1701}.

\bibitem[{\citenamefont{Benesch et~al.}(2014)}]{MOLLER:2014iki}
\bibinfo{author}{\bibfnamefont{J.}~\bibnamefont{Benesch}} \bibnamefont{et~al.}
  (\bibinfo{collaboration}{MOLLER}) (\bibinfo{year}{2014}), \eprint{1411.4088}.

\bibitem[{\citenamefont{Berger et~al.}(2016)}]{Berger:2015aaa}
\bibinfo{author}{\bibfnamefont{N.}~\bibnamefont{Berger}} \bibnamefont{et~al.},
  \bibinfo{journal}{J. Univ. Sci. Tech. China} \textbf{\bibinfo{volume}{46}},
  \bibinfo{pages}{481} (\bibinfo{year}{2016}), \eprint{1511.03934}.

\bibitem[{\citenamefont{Du et~al.}(2021)\citenamefont{Du, Freitas, Patel, and
  Ramsey-Musolf}}]{Du:2019evk}
\bibinfo{author}{\bibfnamefont{Y.}~\bibnamefont{Du}},
  \bibinfo{author}{\bibfnamefont{A.}~\bibnamefont{Freitas}},
  \bibinfo{author}{\bibfnamefont{H.~H.} \bibnamefont{Patel}}, \bibnamefont{and}
  \bibinfo{author}{\bibfnamefont{M.~J.} \bibnamefont{Ramsey-Musolf}},
  \bibinfo{journal}{Phys. Rev. Lett.} \textbf{\bibinfo{volume}{126}},
  \bibinfo{pages}{131801} (\bibinfo{year}{2021}), \eprint{1912.08220}.

\bibitem[{\citenamefont{Bondar et~al.}(2020)\citenamefont{Bondar, Grabovsky,
  Reznichenko, Rudenko, and Vorobyev}}]{Bondar:2019zgm}
\bibinfo{author}{\bibfnamefont{A.}~\bibnamefont{Bondar}},
  \bibinfo{author}{\bibfnamefont{A.}~\bibnamefont{Grabovsky}},
  \bibinfo{author}{\bibfnamefont{A.}~\bibnamefont{Reznichenko}},
  \bibinfo{author}{\bibfnamefont{A.}~\bibnamefont{Rudenko}}, \bibnamefont{and}
  \bibinfo{author}{\bibfnamefont{V.}~\bibnamefont{Vorobyev}},
  \bibinfo{journal}{JHEP} \textbf{\bibinfo{volume}{03}}, \bibinfo{pages}{076}
  (\bibinfo{year}{2020}), \eprint{1912.09760}.

\bibitem[{\citenamefont{Fu et~al.}(2023)\citenamefont{Fu, Li, Wang, Yu, and
  Zhang}}]{Fu:2023ose}
\bibinfo{author}{\bibfnamefont{J.}~\bibnamefont{Fu}},
  \bibinfo{author}{\bibfnamefont{H.-B.} \bibnamefont{Li}},
  \bibinfo{author}{\bibfnamefont{J.-P.} \bibnamefont{Wang}},
  \bibinfo{author}{\bibfnamefont{F.-S.} \bibnamefont{Yu}}, \bibnamefont{and}
  \bibinfo{author}{\bibfnamefont{J.}~\bibnamefont{Zhang}},
  \bibinfo{journal}{Phys. Rev. D} \textbf{\bibinfo{volume}{108}},
  \bibinfo{pages}{L091301} (\bibinfo{year}{2023}), \eprint{2307.04364}.

\bibitem[{\citenamefont{Du et~al.}(2024{\natexlab{b}})\citenamefont{Du, He, Ma,
  and Du}}]{Du:2024jfc}
\bibinfo{author}{\bibfnamefont{Y.}~\bibnamefont{Du}},
  \bibinfo{author}{\bibfnamefont{X.-G.} \bibnamefont{He}},
  \bibinfo{author}{\bibfnamefont{J.-P.} \bibnamefont{Ma}}, \bibnamefont{and}
  \bibinfo{author}{\bibfnamefont{X.-Y.} \bibnamefont{Du}},
  \bibinfo{journal}{Phys. Rev. D} \textbf{\bibinfo{volume}{110}},
  \bibinfo{pages}{076019} (\bibinfo{year}{2024}{\natexlab{b}}),
  \eprint{2405.09625}.

\bibitem[{\citenamefont{Anderle et~al.}(2021)}]{Anderle:2021wcy}
\bibinfo{author}{\bibfnamefont{D.~P.} \bibnamefont{Anderle}}
  \bibnamefont{et~al.}, \bibinfo{journal}{Front. Phys. (Beijing)}
  \textbf{\bibinfo{volume}{16}}, \bibinfo{pages}{64701} (\bibinfo{year}{2021}),
  \eprint{2102.09222}.

\bibitem[{\citenamefont{Liu et~al.}(2003)\citenamefont{Liu, Prezeau, and
  Ramsey-Musolf}}]{Liu:2002bq}
\bibinfo{author}{\bibfnamefont{C.~P.} \bibnamefont{Liu}},
  \bibinfo{author}{\bibfnamefont{G.}~\bibnamefont{Prezeau}}, \bibnamefont{and}
  \bibinfo{author}{\bibfnamefont{M.~J.} \bibnamefont{Ramsey-Musolf}},
  \bibinfo{journal}{Phys. Rev. C} \textbf{\bibinfo{volume}{67}},
  \bibinfo{pages}{035501} (\bibinfo{year}{2003}), \eprint{nucl-th/0212041}.

\bibitem[{\citenamefont{Diaconescu and
  Ramsey-Musolf}(2004)}]{Diaconescu:2004aa}
\bibinfo{author}{\bibfnamefont{L.}~\bibnamefont{Diaconescu}} \bibnamefont{and}
  \bibinfo{author}{\bibfnamefont{M.~J.} \bibnamefont{Ramsey-Musolf}},
  \bibinfo{journal}{Phys. Rev. C} \textbf{\bibinfo{volume}{70}},
  \bibinfo{pages}{054003} (\bibinfo{year}{2004}), \eprint{nucl-th/0405044}.

\bibitem[{\citenamefont{Ramsey-Musolf}(2005)}]{Ramsey-Musolf:2005qlt}
\bibinfo{author}{\bibfnamefont{M.~J.} \bibnamefont{Ramsey-Musolf}},
  \bibinfo{journal}{Eur. Phys. J. A} \textbf{\bibinfo{volume}{24S2}},
  \bibinfo{pages}{197} (\bibinfo{year}{2005}), \eprint{nucl-th/0501023}.

\bibitem[{\citenamefont{Zhao et~al.}(2017)\citenamefont{Zhao, Deshpande, Huang,
  Kumar, and Riordan}}]{Zhao:2016rfu}
\bibinfo{author}{\bibfnamefont{Y.~X.} \bibnamefont{Zhao}},
  \bibinfo{author}{\bibfnamefont{A.}~\bibnamefont{Deshpande}},
  \bibinfo{author}{\bibfnamefont{J.}~\bibnamefont{Huang}},
  \bibinfo{author}{\bibfnamefont{K.~S.} \bibnamefont{Kumar}}, \bibnamefont{and}
  \bibinfo{author}{\bibfnamefont{S.}~\bibnamefont{Riordan}},
  \bibinfo{journal}{Eur. Phys. J. A} \textbf{\bibinfo{volume}{53}},
  \bibinfo{pages}{55} (\bibinfo{year}{2017}), \eprint{1612.06927}.

\bibitem[{\citenamefont{Cirigliano et~al.}(2021)\citenamefont{Cirigliano,
  Fuyuto, Lee, Mereghetti, and Yan}}]{Cirigliano:2021img}
\bibinfo{author}{\bibfnamefont{V.}~\bibnamefont{Cirigliano}},
  \bibinfo{author}{\bibfnamefont{K.}~\bibnamefont{Fuyuto}},
  \bibinfo{author}{\bibfnamefont{C.}~\bibnamefont{Lee}},
  \bibinfo{author}{\bibfnamefont{E.}~\bibnamefont{Mereghetti}},
  \bibnamefont{and} \bibinfo{author}{\bibfnamefont{B.}~\bibnamefont{Yan}},
  \bibinfo{journal}{JHEP} \textbf{\bibinfo{volume}{03}}, \bibinfo{pages}{256}
  (\bibinfo{year}{2021}), \eprint{2102.06176}.

\bibitem[{\citenamefont{Yan et~al.}(2021)\citenamefont{Yan, Yu, and
  Yuan}}]{Yan:2021htf}
\bibinfo{author}{\bibfnamefont{B.}~\bibnamefont{Yan}},
  \bibinfo{author}{\bibfnamefont{Z.}~\bibnamefont{Yu}}, \bibnamefont{and}
  \bibinfo{author}{\bibfnamefont{C.~P.} \bibnamefont{Yuan}},
  \bibinfo{journal}{Phys. Lett. B} \textbf{\bibinfo{volume}{822}},
  \bibinfo{pages}{136697} (\bibinfo{year}{2021}), \eprint{2107.02134}.

\bibitem[{\citenamefont{Liu and Yan}(2023)}]{Liu:2021lan}
\bibinfo{author}{\bibfnamefont{Y.}~\bibnamefont{Liu}} \bibnamefont{and}
  \bibinfo{author}{\bibfnamefont{B.}~\bibnamefont{Yan}},
  \bibinfo{journal}{Chin. Phys. C} \textbf{\bibinfo{volume}{47}},
  \bibinfo{pages}{043113} (\bibinfo{year}{2023}), \eprint{2112.02477}.

\bibitem[{\citenamefont{Li et~al.}(2022)\citenamefont{Li, Yan, and
  Yuan}}]{Li:2021uww}
\bibinfo{author}{\bibfnamefont{H.~T.} \bibnamefont{Li}},
  \bibinfo{author}{\bibfnamefont{B.}~\bibnamefont{Yan}}, \bibnamefont{and}
  \bibinfo{author}{\bibfnamefont{C.~P.} \bibnamefont{Yuan}},
  \bibinfo{journal}{Phys. Lett. B} \textbf{\bibinfo{volume}{833}},
  \bibinfo{pages}{137300} (\bibinfo{year}{2022}), \eprint{2112.07747}.

\bibitem[{\citenamefont{Yan}(2022)}]{Yan:2022npz}
\bibinfo{author}{\bibfnamefont{B.}~\bibnamefont{Yan}}, \bibinfo{journal}{Phys.
  Lett. B} \textbf{\bibinfo{volume}{833}}, \bibinfo{pages}{137384}
  (\bibinfo{year}{2022}), \eprint{2203.01510}.

\bibitem[{\citenamefont{Boughezal et~al.}(2022)\citenamefont{Boughezal, Emmert,
  Kutz, Mantry, Nycz, Petriello, \c{S}im\c{s}ek, Wiegand, and
  Zheng}}]{Boughezal:2022pmb}
\bibinfo{author}{\bibfnamefont{R.}~\bibnamefont{Boughezal}},
  \bibinfo{author}{\bibfnamefont{A.}~\bibnamefont{Emmert}},
  \bibinfo{author}{\bibfnamefont{T.}~\bibnamefont{Kutz}},
  \bibinfo{author}{\bibfnamefont{S.}~\bibnamefont{Mantry}},
  \bibinfo{author}{\bibfnamefont{M.}~\bibnamefont{Nycz}},
  \bibinfo{author}{\bibfnamefont{F.}~\bibnamefont{Petriello}},
  \bibinfo{author}{\bibfnamefont{K.}~\bibnamefont{\c{S}im\c{s}ek}},
  \bibinfo{author}{\bibfnamefont{D.}~\bibnamefont{Wiegand}}, \bibnamefont{and}
  \bibinfo{author}{\bibfnamefont{X.}~\bibnamefont{Zheng}},
  \bibinfo{journal}{Phys. Rev. D} \textbf{\bibinfo{volume}{106}},
  \bibinfo{pages}{016006} (\bibinfo{year}{2022}), \eprint{2204.07557}.

\bibitem[{\citenamefont{Bacchetta et~al.}(2024)\citenamefont{Bacchetta,
  Cerutti, Manna, Radici, and Zheng}}]{Bacchetta:2023hlw}
\bibinfo{author}{\bibfnamefont{A.}~\bibnamefont{Bacchetta}},
  \bibinfo{author}{\bibfnamefont{M.}~\bibnamefont{Cerutti}},
  \bibinfo{author}{\bibfnamefont{L.}~\bibnamefont{Manna}},
  \bibinfo{author}{\bibfnamefont{M.}~\bibnamefont{Radici}}, \bibnamefont{and}
  \bibinfo{author}{\bibfnamefont{X.}~\bibnamefont{Zheng}},
  \bibinfo{journal}{Phys. Lett. B} \textbf{\bibinfo{volume}{849}},
  \bibinfo{pages}{138455} (\bibinfo{year}{2024}), \eprint{2306.04704}.

\bibitem[{\citenamefont{Balkin et~al.}(2024)\citenamefont{Balkin, Hen, Li, Liu,
  Ma, Soreq, and Williams}}]{Balkin:2023gya}
\bibinfo{author}{\bibfnamefont{R.}~\bibnamefont{Balkin}},
  \bibinfo{author}{\bibfnamefont{O.}~\bibnamefont{Hen}},
  \bibinfo{author}{\bibfnamefont{W.}~\bibnamefont{Li}},
  \bibinfo{author}{\bibfnamefont{H.}~\bibnamefont{Liu}},
  \bibinfo{author}{\bibfnamefont{T.}~\bibnamefont{Ma}},
  \bibinfo{author}{\bibfnamefont{Y.}~\bibnamefont{Soreq}}, \bibnamefont{and}
  \bibinfo{author}{\bibfnamefont{M.}~\bibnamefont{Williams}},
  \bibinfo{journal}{JHEP} \textbf{\bibinfo{volume}{02}}, \bibinfo{pages}{123}
  (\bibinfo{year}{2024}), \eprint{2310.08827}.

\bibitem[{\citenamefont{Wang et~al.}(2024)\citenamefont{Wang, Wen, Xing, and
  Yan}}]{Wang:2024zns}
\bibinfo{author}{\bibfnamefont{H.-L.} \bibnamefont{Wang}},
  \bibinfo{author}{\bibfnamefont{X.-K.} \bibnamefont{Wen}},
  \bibinfo{author}{\bibfnamefont{H.}~\bibnamefont{Xing}}, \bibnamefont{and}
  \bibinfo{author}{\bibfnamefont{B.}~\bibnamefont{Yan}},
  \bibinfo{journal}{Phys. Rev. D} \textbf{\bibinfo{volume}{109}},
  \bibinfo{pages}{095025} (\bibinfo{year}{2024}), \eprint{2401.08419}.

\bibitem[{\citenamefont{Wen et~al.}(2024)\citenamefont{Wen, Yan, Yu, and
  Yuan}}]{Wen:2024cfu}
\bibinfo{author}{\bibfnamefont{X.-K.} \bibnamefont{Wen}},
  \bibinfo{author}{\bibfnamefont{B.}~\bibnamefont{Yan}},
  \bibinfo{author}{\bibfnamefont{Z.}~\bibnamefont{Yu}}, \bibnamefont{and}
  \bibinfo{author}{\bibfnamefont{C.~P.} \bibnamefont{Yuan}}
  (\bibinfo{year}{2024}), \eprint{2408.07255}.

\bibitem[{\citenamefont{Deng et~al.}(2025)\citenamefont{Deng, Jiang, Liu, and
  Yan}}]{Deng:2025hio}
\bibinfo{author}{\bibfnamefont{Y.}~\bibnamefont{Deng}},
  \bibinfo{author}{\bibfnamefont{X.-H.} \bibnamefont{Jiang}},
  \bibinfo{author}{\bibfnamefont{T.}~\bibnamefont{Liu}}, \bibnamefont{and}
  \bibinfo{author}{\bibfnamefont{B.}~\bibnamefont{Yan}} (\bibinfo{year}{2025}),
  \eprint{2503.02605}.

\bibitem[{\citenamefont{Blumlein and Kochelev}(1997)}]{Blumlein:1996vs}
\bibinfo{author}{\bibfnamefont{J.}~\bibnamefont{Blumlein}} \bibnamefont{and}
  \bibinfo{author}{\bibfnamefont{N.}~\bibnamefont{Kochelev}},
  \bibinfo{journal}{Nucl. Phys. B} \textbf{\bibinfo{volume}{498}},
  \bibinfo{pages}{285} (\bibinfo{year}{1997}), \eprint{hep-ph/9612318}.

\bibitem[{\citenamefont{Blumlein and Tkabladze}(1999)}]{Blumlein:1998nv}
\bibinfo{author}{\bibfnamefont{J.}~\bibnamefont{Blumlein}} \bibnamefont{and}
  \bibinfo{author}{\bibfnamefont{A.}~\bibnamefont{Tkabladze}},
  \bibinfo{journal}{Nucl. Phys. B} \textbf{\bibinfo{volume}{553}},
  \bibinfo{pages}{427} (\bibinfo{year}{1999}), \eprint{hep-ph/9812478}.

\bibitem[{\citenamefont{Nocera et~al.}(2014)\citenamefont{Nocera, Ball, Forte,
  Ridolfi, and Rojo}}]{Nocera:2014gqa}
\bibinfo{author}{\bibfnamefont{E.~R.} \bibnamefont{Nocera}},
  \bibinfo{author}{\bibfnamefont{R.~D.} \bibnamefont{Ball}},
  \bibinfo{author}{\bibfnamefont{S.}~\bibnamefont{Forte}},
  \bibinfo{author}{\bibfnamefont{G.}~\bibnamefont{Ridolfi}}, \bibnamefont{and}
  \bibinfo{author}{\bibfnamefont{J.}~\bibnamefont{Rojo}}
  (\bibinfo{collaboration}{NNPDF}), \bibinfo{journal}{Nucl. Phys. B}
  \textbf{\bibinfo{volume}{887}}, \bibinfo{pages}{276} (\bibinfo{year}{2014}),
  \eprint{1406.5539}.

\bibitem[{\citenamefont{Harland-Lang et~al.}(2015)\citenamefont{Harland-Lang,
  Martin, Motylinski, and Thorne}}]{Harland-Lang:2014zoa}
\bibinfo{author}{\bibfnamefont{L.~A.} \bibnamefont{Harland-Lang}},
  \bibinfo{author}{\bibfnamefont{A.~D.} \bibnamefont{Martin}},
  \bibinfo{author}{\bibfnamefont{P.}~\bibnamefont{Motylinski}},
  \bibnamefont{and} \bibinfo{author}{\bibfnamefont{R.~S.}
  \bibnamefont{Thorne}}, \bibinfo{journal}{Eur. Phys. J. C}
  \textbf{\bibinfo{volume}{75}}, \bibinfo{pages}{204} (\bibinfo{year}{2015}),
  \eprint{1412.3989}.

\bibitem[{\citenamefont{Martin et~al.}(2009)\citenamefont{Martin, Stirling,
  Thorne, and Watt}}]{Martin:2009iq}
\bibinfo{author}{\bibfnamefont{A.~D.} \bibnamefont{Martin}},
  \bibinfo{author}{\bibfnamefont{W.~J.} \bibnamefont{Stirling}},
  \bibinfo{author}{\bibfnamefont{R.~S.} \bibnamefont{Thorne}},
  \bibnamefont{and} \bibinfo{author}{\bibfnamefont{G.}~\bibnamefont{Watt}},
  \bibinfo{journal}{Eur. Phys. J. C} \textbf{\bibinfo{volume}{63}},
  \bibinfo{pages}{189} (\bibinfo{year}{2009}), \eprint{0901.0002}.

\bibitem[{\citenamefont{Erler and Ramsey-Musolf}(2005)}]{Erler:2004in}
\bibinfo{author}{\bibfnamefont{J.}~\bibnamefont{Erler}} \bibnamefont{and}
  \bibinfo{author}{\bibfnamefont{M.~J.} \bibnamefont{Ramsey-Musolf}},
  \bibinfo{journal}{Phys. Rev. D} \textbf{\bibinfo{volume}{72}},
  \bibinfo{pages}{073003} (\bibinfo{year}{2005}), \eprint{hep-ph/0409169}.

\end{thebibliography}
\end{document}